\documentclass[journal=jpcafh,manuscript=article]{achemso}
\usepackage{longtable}
\usepackage{multirow}
\usepackage{amsmath}
\usepackage{subfigure}
\usepackage[usenames,dvipsnames]{color}
\usepackage{soul}
\usepackage{bm}
\usepackage{rotating}

\usepackage{amssymb}
\usepackage{titlecaps}
\Addlcwords{a an the of on at with for and in into from by without}
\usepackage{soul}

\makeatletter
\newcommand*{\addFileDependency}[1]{
  \typeout{(#1)}
  \@addtofilelist{#1}
  \IfFileExists{#1}{}{\typeout{No file #1.}}
}
\makeatother

\newcommand*{\myexternaldocument}[1]{%
    \externaldocument{#1}%
    \addFileDependency{#1.tex}%
    \addFileDependency{#1.aux}%
}
\usepackage{xr}
\myexternaldocument{si}

\usepackage{multicol} \usepackage{wrapfig} \usepackage{enumitem}
\usepackage{bm} \usepackage{outlines} 
\SectionNumbersOn

\author{Meenu Upadhyay}
\affiliation[University of Basel]{Department of Chemistry, University
  of Basel, Klingelbergstrasse 80 , CH-4056 Basel, Switzerland.}
\author{Markus Meuwly}
\affiliation[University of Basel]{Department of Chemistry, University
  of Basel, Klingelbergstrasse 80 , CH-4056 Basel, Switzerland.}
\email{m.meuwly@unibas.ch}

\title{Energy Redistribution following CO$_2$ Formation on Cold
  Amorphous Solid Water}

\begin{document}

\date{\today}

\begin{abstract}
The formation of molecules in and on amorphous solid water (ASW) as it
occurs in interstellar space releases appreciable amounts of energy
that need to be dissipated to the environment. Here, energy transfer
between CO$_2$ formed within and on the surface of amorphous solid
water (ASW) and the surrounding water is studied. Following CO($^1
\Sigma^+$) + O($^1$D) recombination the average translational and
internal energy of the water molecules increases on the $\sim 10$ ps
time scale by 15 \% to 20 \% depending on whether the reaction takes
place on the surface or in an internal cavity of ASW. Due to tight
coupling between CO$_2$ and the surrounding water molecules the
internal energy exhibits a peak at early times which is present for
recombination on the surface but absent for the process inside
ASW. Energy transfer to the water molecules is characterized by a
rapid $\sim 10$ ps and a considerably slower $\sim 1$ ns
component. Within 50 ps a mostly uniform temperature increase of the
ASW across the entire surface is found. The results suggest that
energy transfer between a molecule formed on and within ASW is
efficient and helps to stabilize the products generated.
\end{abstract}

\section{Introduction}
The motion of adsorbates in and on amorphous solid water (ASW) is
essential for chemistry at astrophysical conditions. Typically, bulk
water is present in the form of ASW which is the main component of
interstellar ices.\cite{hagen:1981} The structure of ASW is usually
probed by spectroscopic measurements\cite{hagen:1981,jenniskens:1994}
although interference-based methods have also been
employed.\cite{linnartz:2012} ASWs are porous structures characterized
by surface roughness and internal cavities of different sizes which
can retain molecular or atomic guests.\cite{barnun:1987} Under
laboratory conditions the water ices seem to be
amorphous\cite{Oba09p464} whereas the morphology of ices in the
interstellar medium are more debated.\cite{keane:2001}\\

\noindent
The high porosity of ASW\cite{bossa:2014,bossa:2015,cazauxs:2015}
makes it a good catalyst for gas-surface reactions involving
oxygen\cite{Ioppolo:2011,romanzin:2011,Chaabouni:2012,oxy.diff.minissale:2013,o2.dulieu:2016,MM.oxy:2018,MM.oxy:2019},
hydrogen\cite{hama:2013},
carbonaceous\cite{co.form.minissale:2013,minissale:2016} and
nitrogen-containing\cite{no1.minissale:2014} species and helps
maintaining those species on or inside
ASW.\cite{dulieu:2016,minissale:2018} This increases the probability
for the reaction partners to diffuse to locations for collisions and
association reactions to occur. As the diffusivity of individual atoms
and small molecules has been established from both, experiments and
simulations,\cite{minissale:2013,MM.oxy:2014,MM.oxy:2018} this is a
likely scenario for formation of molecules on and within ASW.\\

\noindent
As such association reactions are in general exothermic, the energy
released needs to be transferred to environmental degrees of freedom
for the reaction products to stabilize. This is the quest of the
present work which investigates the time scale and degrees of freedom
to receive the energy liberated in the O($^1$D)+CO($^1 \Sigma^+$)
reaction to form ground state CO$_2$($^1 \Sigma_{\rm g}^{+}$). The
chemical precursors for formation of CO$_2$ are believed to be carbon
monoxide and atomic oxygen and the CO+O reaction has been proposed as
a non-energetic pathway, close to conditions in interstellar
environments, for CO$_2$ formation 20 years ago.\cite{roser:2001}
Formation of CO$_2$($^1 \Sigma_{\rm g}^{+}$) from ground state CO($^1
\Sigma^+$) and electronically excited O($^1$D) is barrierless. The
excited atomic oxygen species can, for example, be generated from
photolysis of H$_2$O\cite{klemm:1975} which has a radiative lifetime
of 110 minutes.\cite{garstang:1951}\\

\noindent
Earlier thermoluminescence experiments suggested that the
O($^3$P)+CO($^1 \Sigma^+$) reaction with both reaction partners in
their electronic ground state yields excited CO$_2^*$ which, after
emission of a photon, leads to formation of
CO$_2$.\cite{pimentel:1979} Such a process has also been proposed to
occur on interstellar grains\cite{herbst:2001} and has been confirmed
experimentally\cite{minissale:2013} with an estimated entrance barrier
of 0.014 eV to 0.103 eV for the process on ASW, compared with a value
of 0.3 eV from high-level electronic structure
calculations.\cite{MM.co2:2021} The surrounding water matrix provides
the necessary coupling\cite{roser:2001} to facilitate relaxation of
the $^3$A$'$ or $^3$A$''$ states of CO$_2$ to the $^1$A$'$ ground
state (correlating with linear $^1 \Sigma_{\rm g}^+$).\\

\noindent
For adsorbed species to react on ASW they need to be able to
diffuse. This has been demonstrated from MD simulations with diffusion
coefficients and desorption energies consistent with
experiments.\cite{MM.oxy:2014,chiavassa:2015} Atomic
oxygen\cite{MM.oxy:2018} on ASW experiences diffusional barriers
between $E_{\rm dif} = 0.2$ kcal/mol and 2 kcal/mol (100 K to 1000 K)
compared with values of $E_{\rm dif} = 990_{\rm -360}^{\rm +530}$ K
determined from experiments.\cite{dulieu:2016} For CO, MD simulations
reported\cite{MM.oxy:2019} desorption energies between 3.1 and 4.0
kcal/mol (1560 K to 2012 K or 130 meV to 170 meV), compared with 120
meV from experiments. On non-porous and crystalline water surfaces
submonolayer desorption energies for CO are 1307 K and 1330 K ($\sim
115$ meV), respectively.\cite{noble:2012}\\

\noindent
After recombination O($^1$D)+CO($^1 \Sigma^+$)$\rightarrow$ CO$_2$($^1
\Sigma_{\rm g}^{+}$) the product is in a highly vibrationally excited
state. For it to stabilize, excess internal energy needs to be
channeled into the environment which is the ASW. The present work
characterizes and quantifies energy relaxation of the CO$_2$($^1
\Sigma_{\rm g}^{+}$) product into internal and translational degrees
of freedom of the surrounding water matrix. First, the methods used
are described. Then, results are presented and discussed. Finally,
conclusions are drawn.\\

\section{Computational Methods}
\begin{figure}
 \begin{center}
 \resizebox{0.8\columnwidth}{!}
           {\includegraphics[scale=0.1,clip,angle=0]{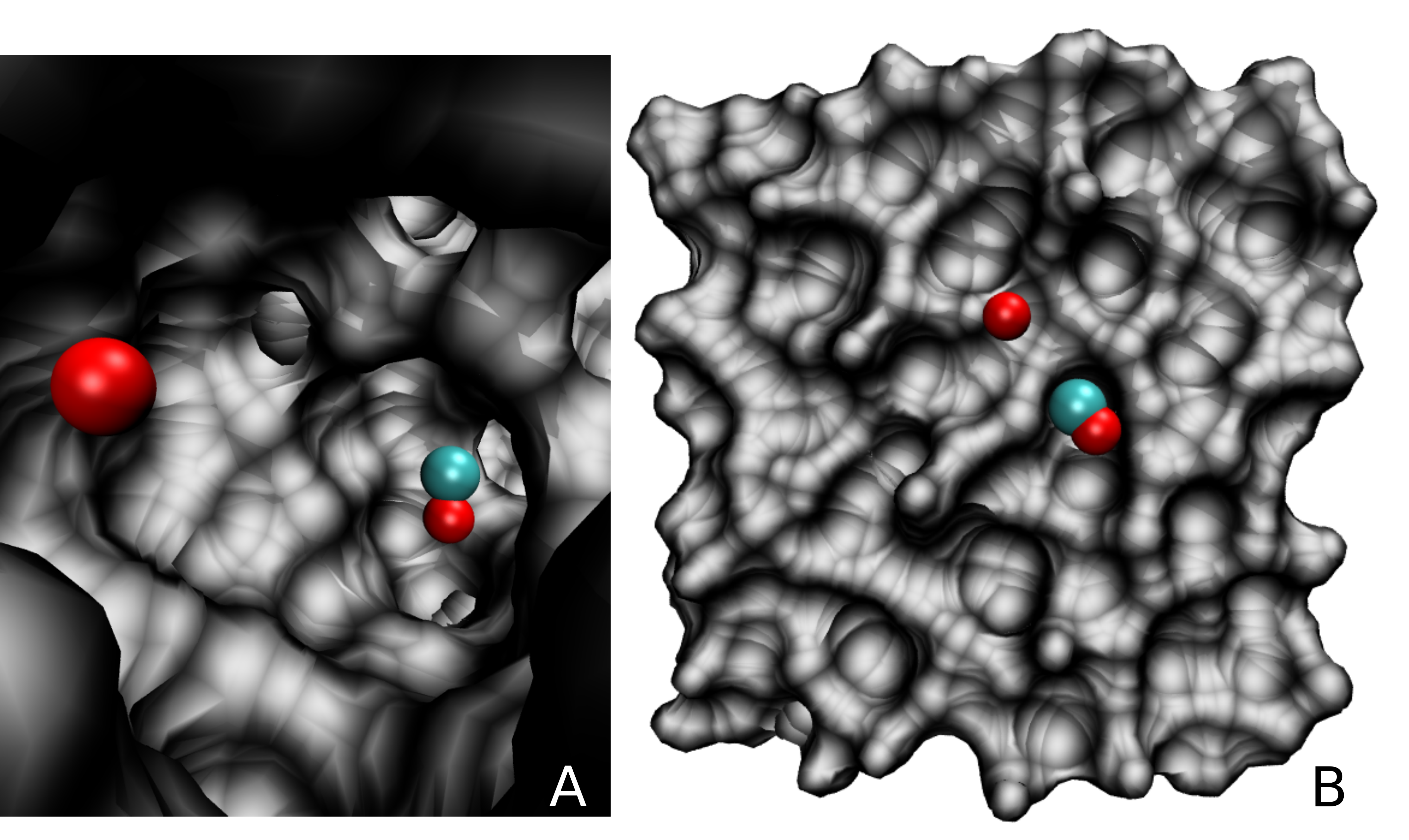}}
           \caption{The simulation system for studying the
             O($^1$D)+CO($^1 \Sigma^+$)$\rightarrow$ CO$_2$($^1
             \Sigma_{\rm g}^{+}$) recombination reaction. Panel A: CO
             and O trapped inside a cavity of ASW; panel B: CO and O
             on the top of the ASW surface.}
\label{fig:fig1}
\end{center}
\end{figure}

All molecular dynamics (MD) simulations were carried out using the
CHARMM suite of programs\cite{charmm.prog} with provisions for bond
forming reactions through multi state adiabatic reactive MD
(MS-ARMD).\cite{msarmd} The simulation system, see Figure
\ref{fig:fig1}, consisted of an equilibrated cubic box of amorphous
solid water with dimension $31 \times 31 \times 31$ \AA\/$^3$
containing 1000 water molecules. As all bonds and angles are flexible,
the simulations were run with a time step of $\Delta t =0.1$ fs and
the non-bonded cutoff was at 13 \AA\/. Simulations were started from
an existing, equilibrated ASW
structure\cite{MM.oxy:2018,MM.oxy:2019,MM.genesis:2021} by adding
CO$_{\rm A}$ and O$_{\rm B}$ inside (Figure \ref{fig:fig1}A) or on top
of (Figure \ref{fig:fig1}B) ASW.\\

\noindent
In the following, the coordinates are the CO stretch $r$, the
separation $R$ between the center of mass of CO$_{\rm A}$ and O$_{\rm
  B}$ and $\theta$ is the O$_{\rm A}$CO$_{\rm B}$ angle. Initial
conditions were generated for a grid of angles $\theta$ and
separations $R$ and simulations were carried out to obtain initial
coordinates and velocities for each of the grid points. With
constrained CO and O position, first 750 steps of steepest descent and
100 steps Adopted Basis Newton-Raphson minimization were carried out,
followed by 50 ps heating dynamics to 50 K. Then, 100 ps equilibration
dynamics was carried out. From each of the runs coordinates and
velocities were saved regularly to obtain initial conditions for each
combination of angle and distance. Production simulations 500 ps or 6
ns in length were then run from saved coordinates and velocities in
the $NVE$ ensemble. Data (energies, coordinates and velocities) were
saved every 1000 steps for subsequent analysis.\\

\noindent
Water was described by a
reparametrized,\cite{Burnham97p6192,MM.ice:2008} flexible (Kumagai,
Kawamura, Yokokawa - KKY) model\cite{kky_orig} was used. The typical
water modes that couple in the $\sim 2000$ cm$^{-1}$ region relevant
in the present work are the water bend (1600 cm$^{-1}$) and the
framework rotation (600 cm$^{-1}$) as was also found for the
vibrational relaxation of cyanide in water.\cite{MM.cn:2011} To
describe CO$_{\rm A}$+O$_{\rm B}$ recombination to form CO$_2$ the
Morse-Morse-Harmonic (MMH) parametrization was
employed.\cite{MM.genesis:2021} This model treats the two CO bonds
with a Morse potential and the OCO bend as a harmonic function. MMH is
a computationally efficient model (fitted to MRCI/aug-cc-pVTZ data),
which yields results for recombination probabilities on ASW comparable
to a more elaborate reproducing kernel Hilbert space (RKHS)
representation.\cite{MM.genesis:2021,MM.co2:2021}\\

\noindent
For CO$_2$, the partial charges were $q_{\rm O} = -0.3e$ and $q_{\rm
  C} = 0.6e$ with standard van der Waals parameters from CHARMM. These
charges are consistent with those obtained from B3LYP/6-31G(d,p)
calculations snapshots from the MD simulations with CO$_2$ adsorbed to
a small water cluster (H$_2$O)$_{10}$ which yield $q_{\rm C} = 0.73e$
and $q_{\rm O} = -0.35e$. This compares with charges of $q_{\rm C} =
0.22e$ and $q_{\rm O} = -0.21e$ for the CO molecule and $q_{\rm O} =
-0.1e$ for an oxygen atom adsorbed to (H$_2$O)$_{10}$. To assess the
dependence of the results on the partial charges used, additional
reactive MD simulations using the MMH parametrization were carried out
with $q_{\rm O} = -0.1e$ and $q_{\rm C} = 0.2e$ (i.e. $q_{\rm CO} =
0.1e$) and with $q_{\rm O} = -0.2e$ and $q_{\rm C} = 0.4e$
(i.e. $q_{\rm CO} = 0.2e$). In all cases, recombination was found to
speed up compared with $q_{\rm CO} = 0.3e$ and $q_{\rm O} = -0.3e$ due
to the increased mobility of the CO molecule and the O atom on the ASW
when reduced partial charges are used.\\

\noindent
The main focus of the present work is to study the energy
redistribution within the system following recombination of CO$_{\rm
  A}$ + O$_{\rm B}$ to form CO$_2$. For this, the average total,
translational and internal energy of the water molecules is analyzed
for recombination on top of and inside ASW. Both, the time scale and
amount of energy dissipated into translational and internal degrees of
freedom was determined.\\

\section{Results and Discussion}
In the following, the energy distribution in the water matrix of the
ASW is separately discussed on the $\sim 100$ ps and on the nanosecond
time scale. Next, the energy flow away from the recombination site is
analyzed and, finally, the energy redistribution to neighboring water
molecules surrounding the recombination site is considered.\\

\subsection{Recombination on the 100 ps time scale}
\begin{figure}
 \begin{center}
 \resizebox{1.0\columnwidth}{!}
           {\includegraphics[scale=0.1,clip,angle=0]{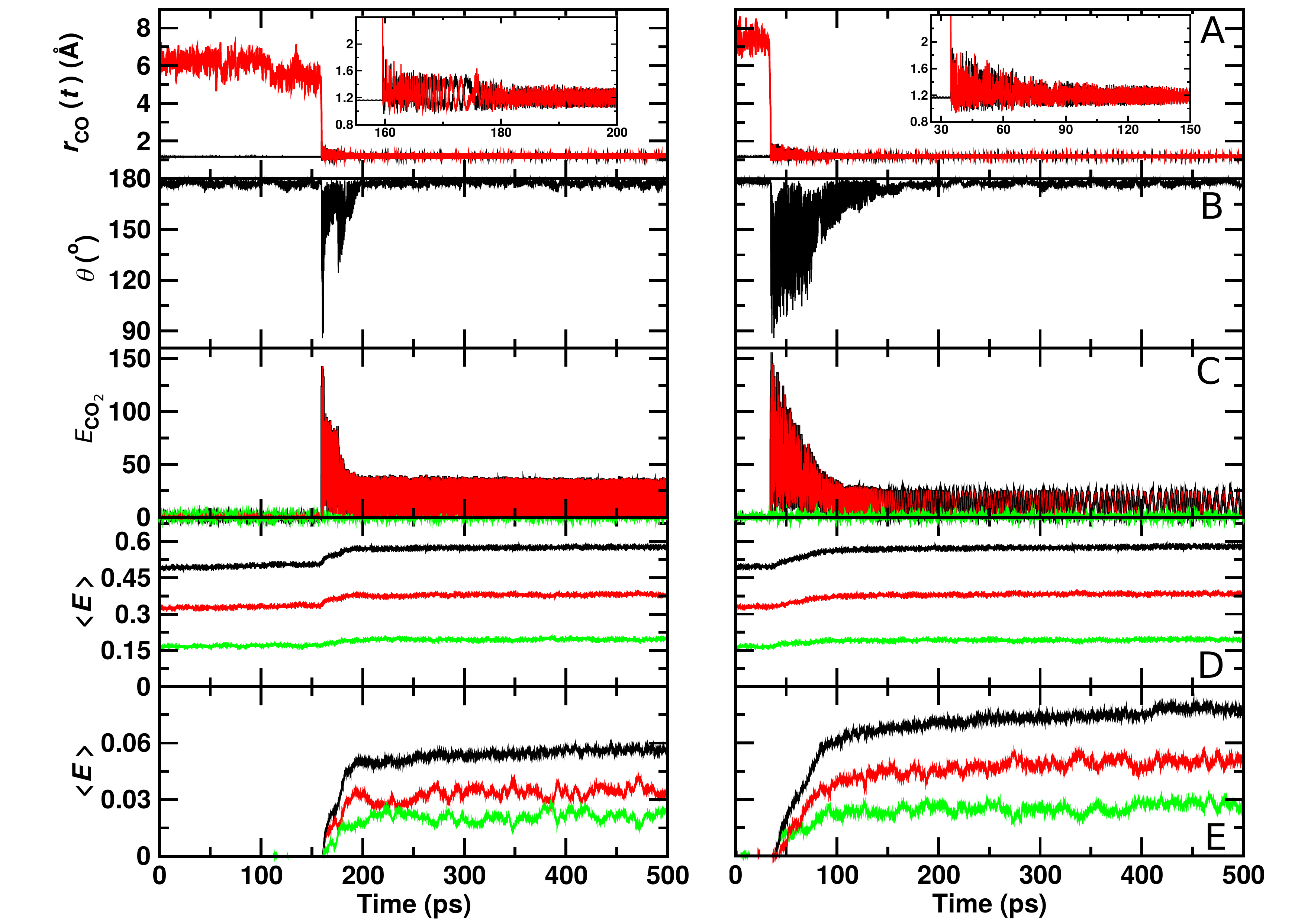}}
 \caption{Recombination of O($^1$D)+CO($^1 \Sigma^+$) to form ground
   state CO$_2 (^1 \Sigma_{\rm g})$ in the ASW cavity (left column)
   and on top of the ASW surface (right column). Panel A: O$_{\rm
     B}$--C$_{\rm CO}$ separation (red) and CO$_{\rm A}$ separation
   (black); Panel B: the O-C-O angle $\theta$; Panels C to E: the
   average total (black), internal (red), and translational (green)
   energies for the CO$_2$ molecule (panel C), the average per water
   molecule (panel D), and the magnitude of the average per water
   molecule relative to the energy before recombination (panel E).}
\label{fig:fig2}
\end{center}
\end{figure}

A typical trajectory for CO$_{\rm A}$+O$_{\rm B}$ recombination inside
the ASW cavity is shown in Figure \ref{fig:fig2} (left
column). Initially, the C--O$_{\rm B}$ separation is $\sim 6$ \AA\/
(Figure \ref{fig:fig2}A). Within 150 ps recombination takes place and
angular distortions lead to exploration of angles $\theta \sim
90^\circ$ (Figure \ref{fig:fig2}B). Relaxation of the angle occurs
within the following 50 ps and the CO$_2$ molecule remains in an
internally excited state on much longer time scales, see Figure
\ref{fig:fig2}C.\cite{MM.genesis:2021} Concomitantly, the average
internal energy of the surrounding water molecules increases by about
10 \%, see black, red and green traces in panels D and E of Figure
\ref{fig:fig2}. The translational (phononic) modes (green) acquire
approximately 1/3 of the additional energy whereas the internal energy
(red) increases by the remaining 2/3.\\

\noindent
Figure \ref{fig:fig2} (right column) reports a recombination
trajectory on top of ASW. In this case, recombination takes place
after $\sim 35$ ps and wide angular excursions extend out to 100
ps. The amount of energy picked up by the water matrix is larger
compared to recombination inside ASW (Panels D and E in figure
\ref{fig:fig2}). The average total energy per water molecule increases
by close to 20 \% and the amount that goes into internal degrees of
freedom is considerably larger. For the translational modes, the
energy after recombination is comparable to that for recombination
within the cavity.\\

\begin{figure}
 \begin{center}
 \resizebox{0.8\columnwidth}{!}
{\includegraphics[scale=0.1,clip,angle=0]{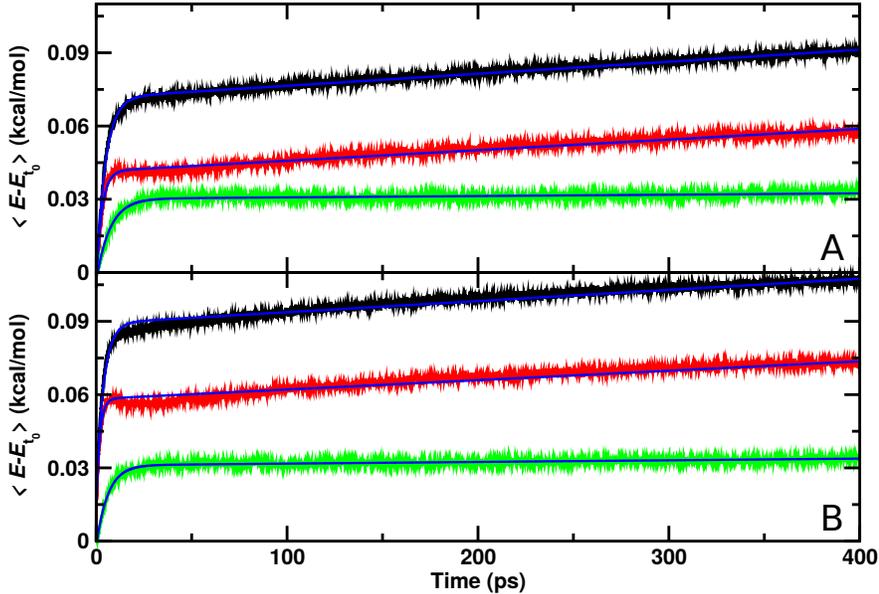}}
\caption{Average total, internal and translational energies for water
  over 70 independent runs relative to the average before
  recombination. The time of reaction for all trajectories is shifted
  to $t=0$ and defined by the first instance at which $r_{\rm C-O_{\rm
      B}} < 1.6$. Panel A: recombination within the ASW cavity. Panel
  B: recombination for CO$_{\rm A}$ + O$_{\rm B}$ on the top of the
  ASW surface. The blue solid line is a fit to an empirical expression
  $\epsilon = a_0e^{-t/a_1}+a_2t + a_3$, see text.}
\label{fig:fig3}
\end{center}
\end{figure}

\noindent
From a set of 70 recombination trajectories for the reaction within
the cavity and on top of the ASW surface, the averaged energy contents
in translational, internal and all degrees of freedom of the water
molecules were determined (see Figure \ref{fig:fig3}). For this
analysis, the time of reaction was set to zero $(t=0)$ to align all
reactive trajectories and all energies are reported relative to the
averages before recombination. The translational contribution for
recombination within and on top of ASW re-equilibrates on the $\sim
25$ ps time scale after which no change in the phononic degrees of
freedom is observed. Contrary to that, the internal degrees of freedom
(red traces) show temporal evolution on two time scales: a rapid phase
on the picosecond time scale, followed by a slow, long increase in the
internal energies. This is also reflected in the averaged total energy
(black).\\

\noindent
As for the single trajectories, the amount of energy released from the
recombination reaction into the translational degrees of freedom is
similar for the reaction inside the cavity and on top of the ASW
surface. For the internal degrees of freedom, however, recombination
on top of the ASW surface leads on an average increase per water
molecule by 0.075 kcal/mol within 400 ps (Figure \ref{fig:fig3}B)
compared with 0.06 kcal/mol for the process inside the cavity. Also,
there is a characteristic decrease in the internal contribution for
recombination on the surface after 15 ps which is even present when
averaging over 70 independent runs. This feature is not found for
recombination within ASW.\\

\noindent
To estimate approximate time scales for the different processes
involved, the average energies were fitted to an empirical expression
$\epsilon = a_0 e^{-t / a_1} + a_2 t +a_3$ where $\epsilon$ is any of
the energies considered. Such a functional form was chosen after
inspection of the data in Figure \ref{fig:fig3} and accounts for the
rapid initial increase in the three energies together with the slow
variation of the internal energy on longer times. This parametrization
is not able to model the dip around 15 ps for recombination on to of
the surface, though. The time scales $a_1$ for total, internal, and
translational energies are [4.8, 2.9, 7.1] ps for recombination inside
the cavity and speed up to [3.9, 1.9, 6.1] ps for the process on the
ASW surface. It is of interest to note that the rapid time scale for
the internal energy is considerably faster than the kinetics of the
translational degrees of freedom for both types of recombinations. The
parameter $a_2$ which describes the slow increase of internal energy
has a value of $a_2 = 4.3 \times 10^{-2}$ (kcal/mol)/ns for
recombination in the cavity and $a_2 = 3.8 \times 10^{-2}$
(kcal/mol)/ns for the reaction on the surface, and is vanishingly
small for the translational energy.\\

\noindent
Average internal energies from representative independent runs for
recombination inside the cavity and on top of the ASW surface are
shown in Figures S1 and S2. For
recombination inside the cavity (Figure S1) the results
confirm that the energy content in the internal degrees of freedom
increases considerably faster than for the translation. Also, it is
found that the amount of energy transferred to translation after
recombination is smaller than that partitioned into internal degrees
of freedom. For recombination on the ASW surface the same observations
are made. In addition, the pronounced maximum after $\sim 5$ ps is
present in all examples shown in Figure S2. To provide a
molecularly resolved interpretation of this feature the HOH angle time
series $\theta(t)$ was analyzed for a trajectory in which CO+O
recombination occurred after 35 ps, see Figure S3. At
the time of reaction the water bending angle decreases from its
average equilibrium value by $\langle \Delta \theta \rangle \sim
1^\circ$ over the next 70 ps after which it relaxes back to the
original value. The signature in the internal energy extends over
$\sim 30$ ps, see Figure S2. Hence, it is possible that
changes in the average water geometry following CO+O recombination are
responsible for the overshooting and subsequent relaxation of the
internal energy for the reaction on the surface. Contrary to that,
recombination within the cavity is less constrained by the direct
interaction with the water molecules which apparently prevents this
particular signature in the internal energy to occur.\\

\subsection{Recombination Dynamics on Longer Time Scales}
\begin{figure}
 \begin{center}
 \resizebox{0.8\columnwidth}{!}
           {\includegraphics[scale=0.1,clip,angle=0]{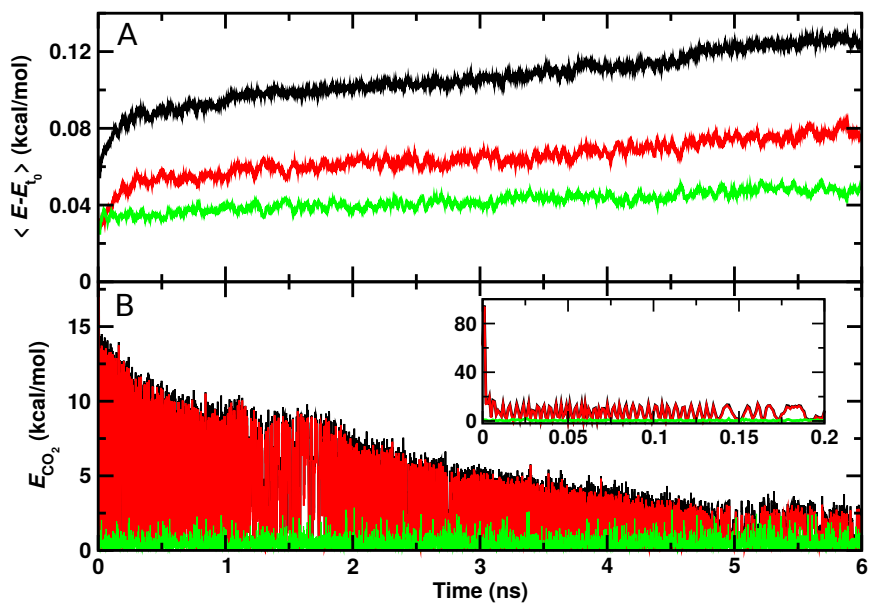}}
\caption{Total (black), internal (red), and translational (green)
  energies for water molecules (panel A) and for the recombined CO$_2$
  molecule (panel B) form a 6 ns rebinding trajectory on the top of
  the ASW surface. The time of reaction is shifted to $t = 0$ and
  defined by the first instance at which $r_{\rm C-O_{\rm B}} < 1.6$
  \AA\/. CO$_2$ continues to relax and the energy in the ASW further
  increases beyond the maximum simulation time of 6 ns after CO$_2$
  recombination.}
\label{fig:fig6}
\end{center}
\end{figure}

It is also of interest to analyze the energy redistribution on the
multi-nanosecond time scale. Figure \ref{fig:fig6}A demonstrates that
the average total kinetic energy per water molecule continuously
increases even on the nanosecond time scale. Most of this increase is
due to the internal degrees of freedom although the translational
component also shows a continuous slow increase on the nanosecond time
scale.\\

\noindent
The relaxation of the CO$_2$ internal energy is reported in Figure
\ref{fig:fig6}B. Within the first few picoseconds (inset) the internal
energy is quenched to $\sim 10$ kcal/mol after which two relaxations
are observed. A first phase during 1 nanosecond following
recombination and a second, slower phase extending out to 6 ns and
beyond. By the end of the simulation the average internal energy of
the CO$_2$ molecule has decreased to $\sim 2.5$ kcal/mol on
average. Hence, it is expected that energy transfer to the surrounding
water continues but slows down considerably on the 10 ns time scale
and longer.\\

\subsection{Energy Migration around the Recombination Site}
\begin{figure}
 \begin{center}
 \resizebox{0.99\columnwidth}{!}
           {\includegraphics[scale=0.2,clip,angle=0]{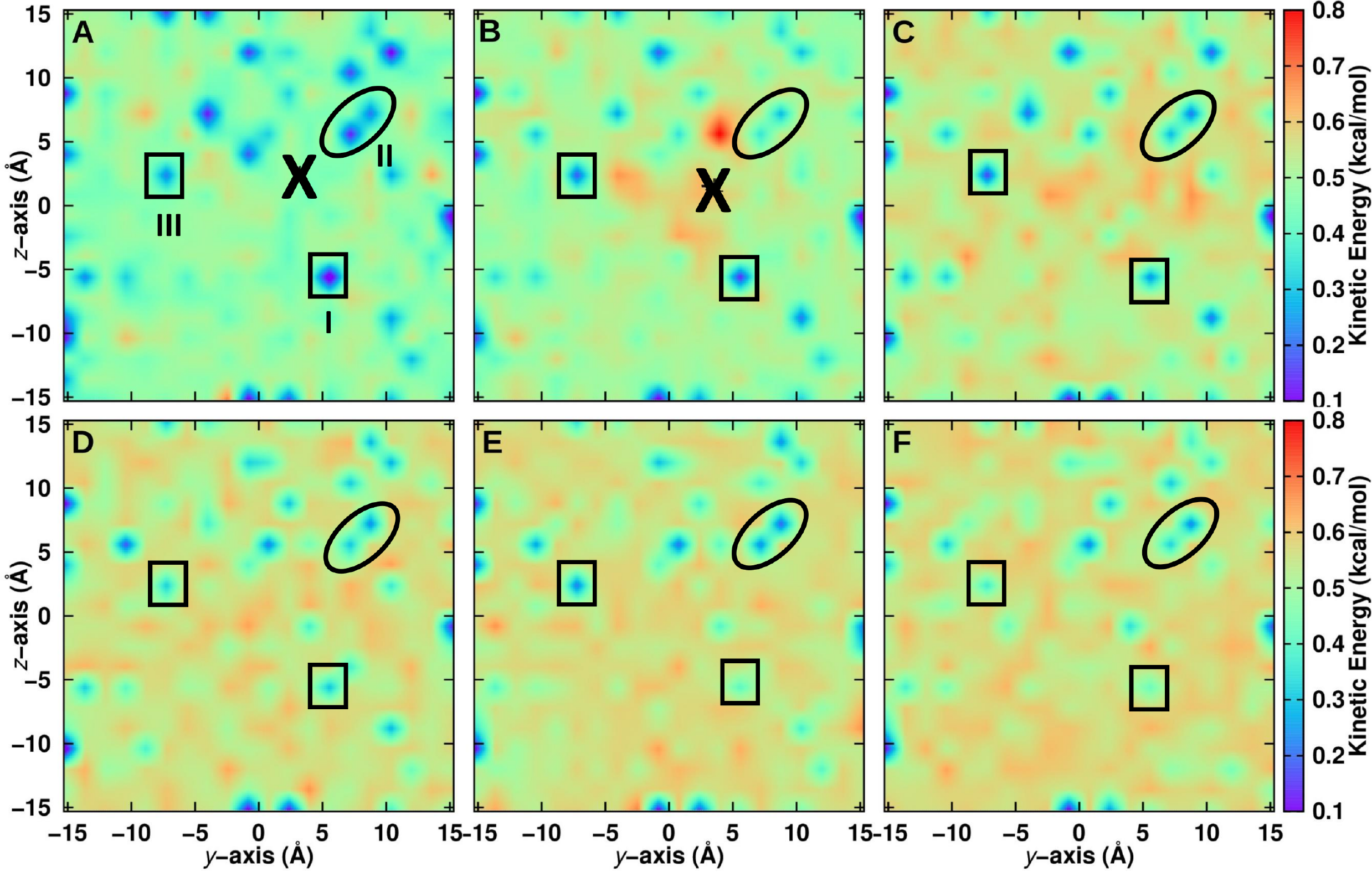}}
 \caption{Kinetic energy of water molecules projected onto the
   $(y,z)-$plane averaged over 7 independent simulations on the top
   layer of the ASW surface. Recombination of CO and O takes place at
   the location labelled with ``X''. Before recombination (Panel A:
   $-5 \leq t \leq 0$ ps) and after recombination (Panel B: $0 \leq t
   \leq 5$ ps, Panel C: $5 \leq t \leq 10$ ps, Panel D: $10 \leq t
   \leq 50$ ps, Panel E: $50 \leq t \leq 100$ ps and Panel F: $100
   \leq t \leq 200$ ps). The average position of CO is indicated by a
   large black cross. Noteworthy regions are labelled I to III and
   surrounded by solid lines. Region I is cool at early times and
   gradually warms up. Region II remains cool for most of the
   simulation time and region III alternates between cool and
   warm. For results within 10 \AA\/ of the surface, see Figure
   S4.}
    \label{fig:fig7}
\end{center}
\end{figure}

For a positionally resolved picture of energy flow the simulation
system was separated in voxels with dimension $31 \times 1 \times 1$
\AA\/$^3$. The kinetic energy of all water molecules within one such
voxel was averaged along the trajectory and projected onto the
$(y,z)-$plane. Which water molecules belong to a particular voxel was
decided based on the water-oxygen atom coordinates. Figure
\ref{fig:fig7}A reports the distribution of total kinetic energy
distribution before recombination. The recombination site is at $(y =
2, z = 2)$ \AA\/ and marked as a large cross. Within the first 5 ps
after recombination the kinetic energy of water molecules within $\sim
10$ \AA\/ of the recombination site increases considerably, by up to a
factor of 4. Following this, energy redistributes continuously across
the entire surface on the 200 ps time scale, see panels C to E.\\

\noindent
Certain regions that are initially ``cold'' (blue) - e.g. the region
labelled ``I'' at $(y=5, z=-5)$ \AA\/ in Figure \ref{fig:fig7} - warm
up as energy transfer from CO$_2$ to the water molecules
occurs. Conversely, other regions remain ``cool'', such as region
``II'' around $(y=5, z=5)$ \AA\/ for which the color code remains blue
until 200 ps. Yet for other regions, such as ``III'', the total
kinetic energy oscillates between cooler and warmer. It is also
instructive to include only the first few ASW layers in this analysis
which was done in Figure S4. Here, the voxels have sizes
$10 \times 1 \times 1$ \AA\/. For one, the cool regions are more
extended before recombination. After recombination energy transfer
occurs in a similar fashion as for the full system. However, the warm
regions are less extended. This suggests that energy transfer also
occurs to a considerable extent {\it into} the bulk rather than across
the surface of the ASW even for recombination on top of ASW.\\

\subsection{Energy Flow to Nearby Water Molecules}
\begin{figure}
\begin{center}
\resizebox{0.8\columnwidth}{!}
          {\includegraphics[scale=0.1,clip,angle=0]{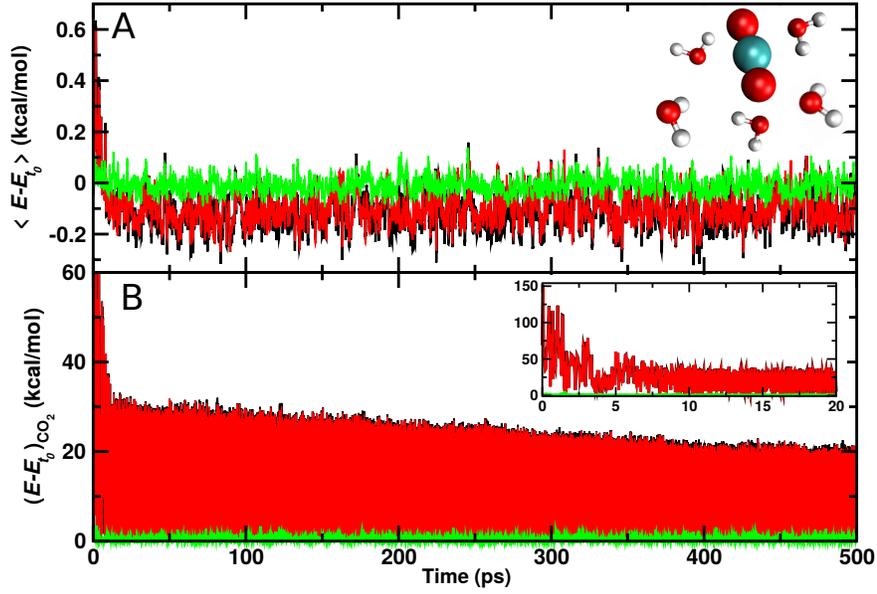}}
\caption{Average total (black), internal (red), and translational
  (green) energies for 5 water molecules (panel A) closest to the
  CO$_2$, and the CO$_2$ molecule (panel B) formed from a
  recombination trajectory on the ASW surface. The time of reaction is
  shifted to $t = 0$ and defined by the first instance at which
  $r_{\rm C-O_{\rm B}} < 1.6$. The initial time scale for energy
  redistribution from the relaxing CO$_2$ molecule to internal and
  translational degrees of freedom of the ASW occurs on the 10 ps time
  scale with slow gradual relaxation on the $\sim 100$ ps to ns time
  scale.}
\label{fig:fig8}
\end{center}
\end{figure}

Finally, individual water molecules in immediate proximity of the
recombination site are analyzed. For one trajectory with recombination
on the ASW surface the average total, internal, and translational
energies for the 5 water molecules closest to the recombination site
are reported in Figure \ref{fig:fig8}. During the first 10 ps after
recombination the average total kinetic energy increases by up to 0.6
kcal/mol per water molecule. Conversely, the translational energy
contribution fluctuates around zero which indicates that the local
structure of ASW remains intact and most of the energy flows into
internal degrees of freedom.\\

\noindent
After this initial increase, cooling takes place with a long-time
average of $-0.1$ kcal/mol per water molecule in the internal degrees
of freedom. On the 500 ps no noticeable changes in the translational
energy content is observed. For the CO$_2$ molecule (see Figure
\ref{fig:fig8}B) the translational energy remains small throughout the
trajectory whereas the internal energy decreases rapidly within the
first 5 ps following recombination. Subsequently, slow cooling on the
100 ps to nanosecond time scale takes place as was already found
earlier, see Figure \ref{fig:fig6}.\\

\section{Discussion and Conclusion}
The present work reports on the energy redistribution across ASW
following O($^1$D)+CO($^1 \Sigma^+$) recombination to form CO$_2 (^1
\Sigma_{\rm g}^+)$ on the surface and in a cavity. It is found that
energy distribution occurs in two phases, one on the picosecond and
one on the nanosecond time scale for both locations. Although the time
dependence of the processes is similar for the two different
recombination sites (inside vs. on top), the dynamics differs in a
number of ways. Firstly, recombination on the surface leads to excess
internal energy on the picosecond time scale which subsequently
relaxes and additional energy transfer into water modes occurs on
longer time scales. Secondly, recombination within the cavity
considered here leads to smaller magnitude $(\sim 15 \%)$ of energy
transferred per water molecule compared with the process on the
surface $(\sim 25 \%)$. Energy relaxation of the CO$_2$ from
recombination on the surface extends over longer times than for the
process in the cavity. Finally, heating of the water molecules occurs
on the 10 ps time scale following the recombination reaction.\\

\noindent
It is of interest to note that - ultimately - energy redistribution in
such systems follows quantum mechanical principles. The present
results suggest that the local energy generated from CO+O
recombination is probably sufficient to excite internal modes of
individual water molecules surrounding the recombination site. Hence,
after CO+O recombination the ASW will be in a state characterized by a
few internally and vibrationally excited water molecules embedded into
a matrix of water molecules in the ground state. Earlier work on a
related problem - the vibrational relaxation of a quantum oscillator
coupled to oscillators of a biomolecule\cite{stock:2009} - found that
using classical mechanics leads to qualitatively correct results
compared with a full quantum treatment. For the relaxation times a
moderate factor of 2 for the difference between classical and rigorous
quantum simulations was reported. Hence, for the present problem it is
also expected that similar conclusions apply and that the
nonequilibrium relaxation dynamics of individual vibrationally excited
water molecules surrounded by vibrationally cold water molecules can
be captured qualitatively from using classical dynamics.\\

\noindent
In summary, the present work demonstrates that O($^1$D)+CO($^1
\Sigma^+$) recombination to form CO$_2 (^1 \Sigma_{\rm g}^+)$ leads to
excitation of both, phononic and internal modes of the water molecules
that constitute the ASW. The time scales for this are on the pico- and
nano-second and lead to warming the water matrix. Water molecules in
direct proximity of the recombination site may become vibrationally
excited and the time scale for their relaxation back to the ground
state will depend on the coupling to the immediate environment. Full
relaxation of the CO$_2$ molecule is expected to occur on the several
10 to 100 nanosecond time scale.\\

\section*{Acknowledgments}

\noindent
The authors gratefully acknowledge financial support from the Swiss
National Science Foundation through grant 200021-117810 and to the
NCCR-MUST.\\

\bibliography{astro3}

\end{document}


\begin{figure}
\begin{center}
\resizebox{0.8\columnwidth}{!}
{\includegraphics[scale=0.1,clip,angle=0]{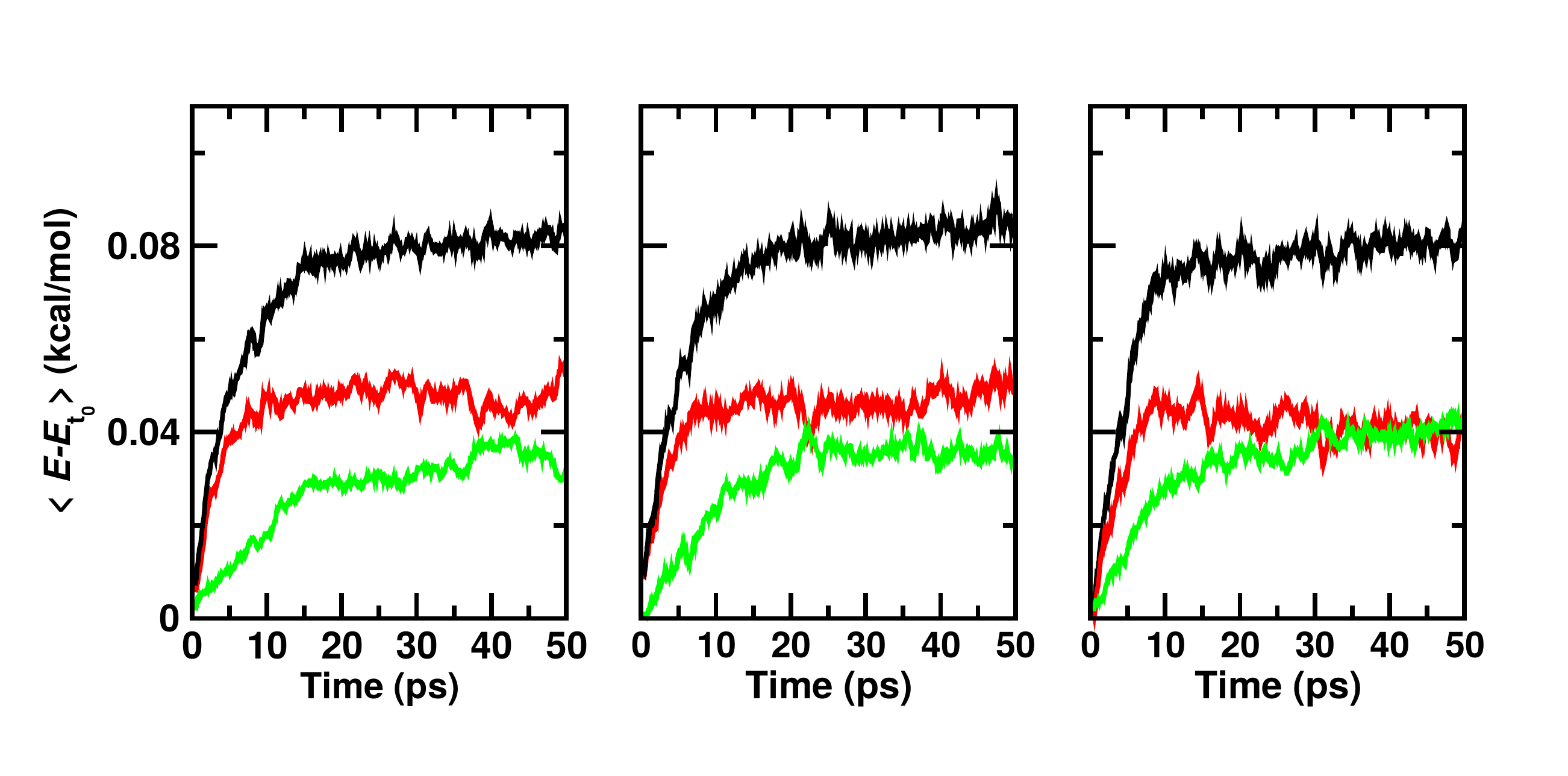}}
\caption{Total (black), internal (red), and translational (green)
  energies for water molecules from 3 independent simulations inside
  the ASW cavity. The time of reaction for all is shifted to $t = 0$
  and defined by the first instance at which $r_{\rm C-O_{\rm B}} <
  1.6$.}
\label{sifig:fig4}
\end{center}
\end{figure}

\begin{figure}
 \begin{center}
 \resizebox{0.8\columnwidth}{!}
 {\includegraphics[scale=0.1,clip,angle=0]{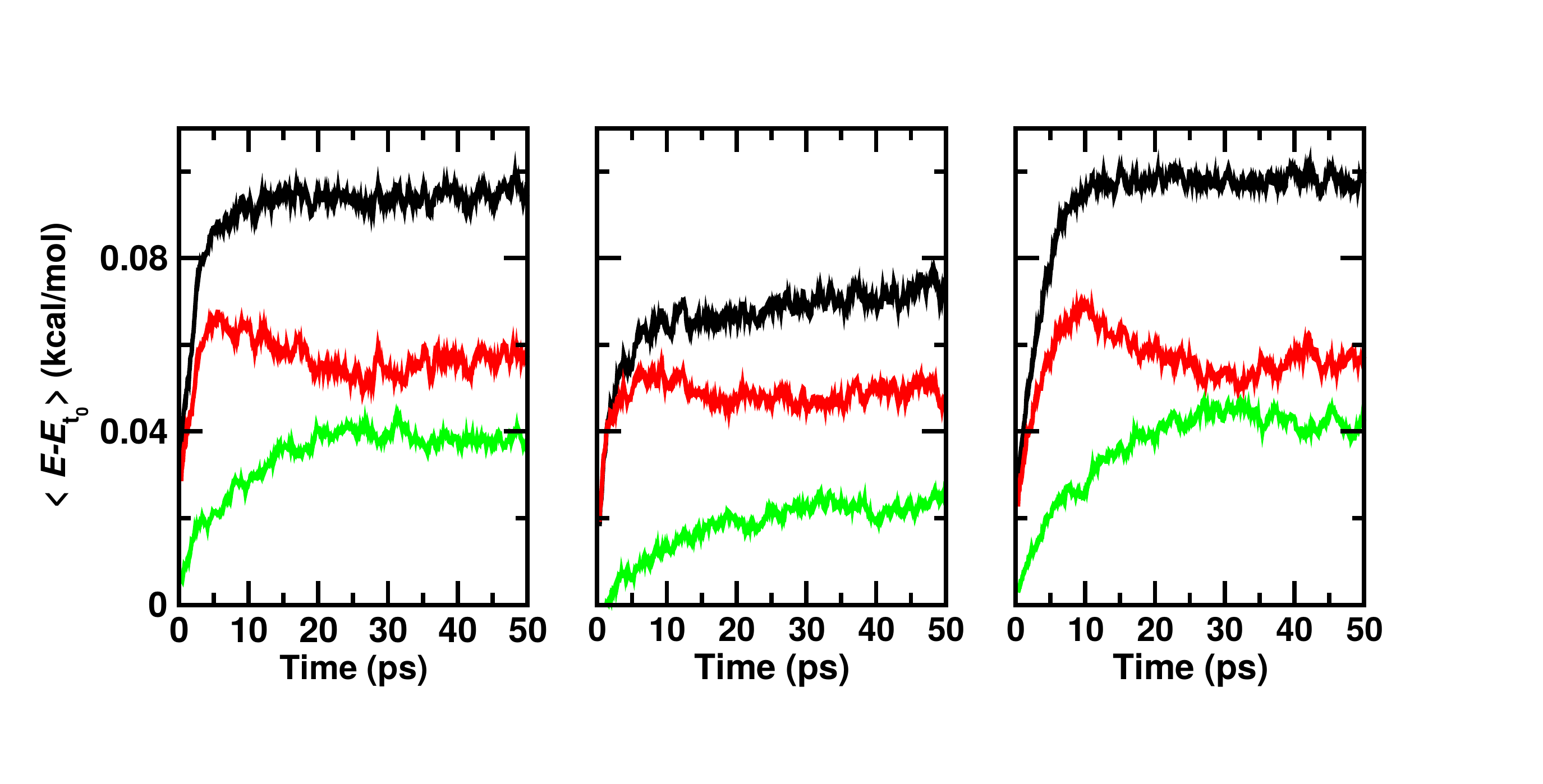}}
 \caption{Total (black), internal (red), and translational (green)
   energies for water molecules from 3 independent simulations on the
   top of the ASW surface. The time of reaction for all is shifted to
   $t = 0$ and defined by the first instance at which $r_{\rm C-O_{\rm
       B}} < 1.6$.}
\label{sifig:fig5}
\end{center}
\end{figure}

\begin{figure}[H]
\centering \includegraphics[scale=0.64]{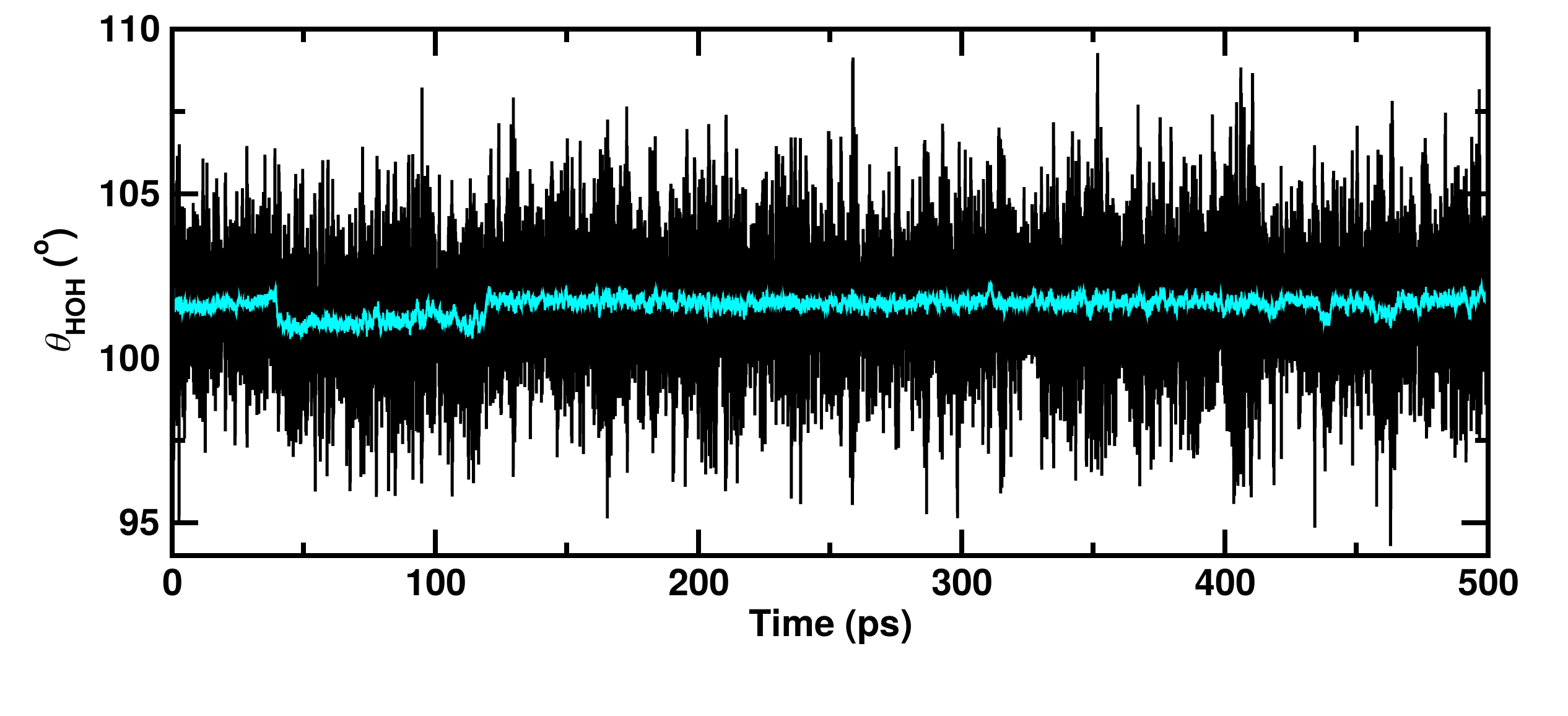}
\caption{Water bending angle time series for a trajectory with
  recombination on the surface at $t = 35$ ps.}
\label{sifig:fig1}
\end{figure}

\begin{figure}[H]
  \centering \includegraphics[scale=0.64]{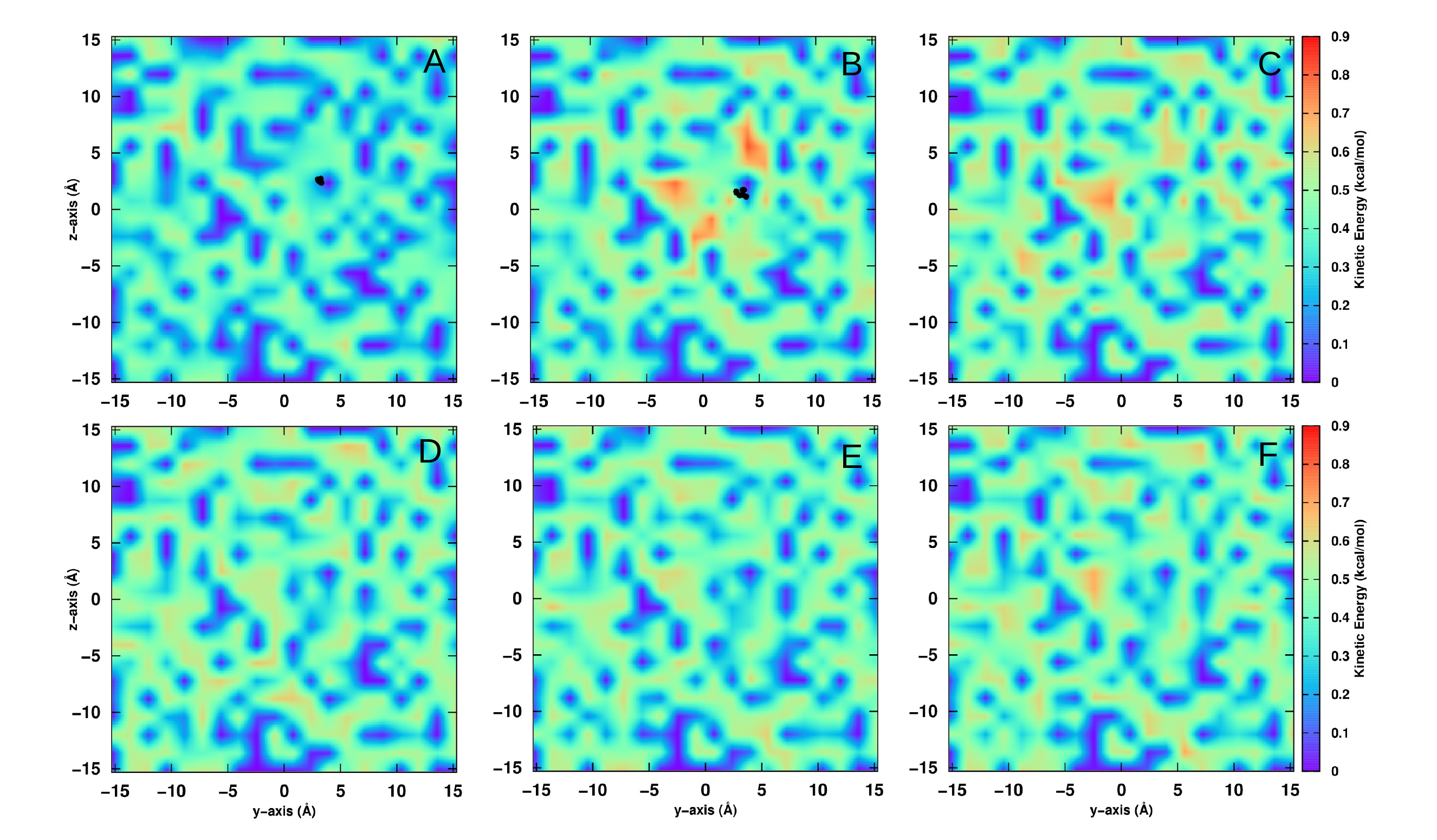}
\caption{Kinetic energy of water molecules projected onto the
  $(y,z)-$plane averaged over 7 independent simulations on the top
  layer of ASW surface. The kinetic energy is averaged over voxels $10
  \times 1 \times 1$ \AA\/$^3$, i.e. only within 10 \AA\/ of the
  surface. Here, $t = 0$ is defined as the first instance at which
  $r_{\rm C-O_{\rm B}} < 1.6$ and all times are aligned with respect
  to this reference. Before recombination (Panel A: $-5 \leq t \leq 0$
  ps) and after recombination (Panel B: $0 \leq t \leq 5$ ps, Panel C:
  $5 \leq t \leq 10$ ps, Panel D: $10 \leq t \leq 50$ ps, Panel E: $50
  \leq t \leq 100$ ps and Panel F: $100 \leq t \leq 200$ ps). The
  average position of CO is indicated by black filled circles.}
\label{sifig:fig2}
\end{figure}
